\documentclass[review,12pt]{elsarticle}

\usepackage{hyperref}

\journal{Journal of Engineering Structures}

\usepackage[margin=1.in]{geometry}
\usepackage{setspace}

\usepackage{hyperref}
\hypersetup{
    colorlinks = true,
     }
\usepackage{amsmath}
\usepackage{amssymb}

\biboptions{sort&compress}
\usepackage{sans}
\usepackage{newtxmath}

\usepackage{caption}
\begin{document}

\begin{frontmatter}

\title{Overview and Analysis of Seismic Resilient Structures with Modular Rocking Shear Walls}
\author{Mehrdad Aghagholizadeh\corref{cor1}}
\cortext[cor1]{Corresponding author}
\ead{mehrdad.aghagholizadeh@lmu.edu}
\ead[link]{https://sehm.lmu.build/}
\address{Assistant Professor, Department of Civil and Environmental Engineering, Loyola Marymount University\\Los Angeles, CA 90045}


\begin{abstract}

The high occupancy rates in urban multi-story buildings, combined with present safety concerns, necessarily prompt a reassessment of performance goals. Given the notable seismic damage and instances of weak-story failures that have been documented after major earthquakes, this paper studies the use of modular shear walls that are free to rock above their foundation. This paper first provides a comprehensive background in analysis of rocking elements such as columns and shear-walls. Then discusses different configurations of rocking-shear-walls. Next, the paper provides two numerical case studies on 9-story and 20-story moment-resisting frames using OpenSees. The floor displacement and interstory drifts under various earthquake excitations for both structures compared for the cases of with and without modular rocking walls. The result shows that the addition of rocking-shear-wall, makes the first mode of the frame becomes dominant which enforces a uniform distribution of interstory drifts that would avoid a soft-story failure.

\end{abstract}


\end{frontmatter}

\section{Introduction}
\begin{figure}[t]
\centering
\includegraphics[width=\textwidth]{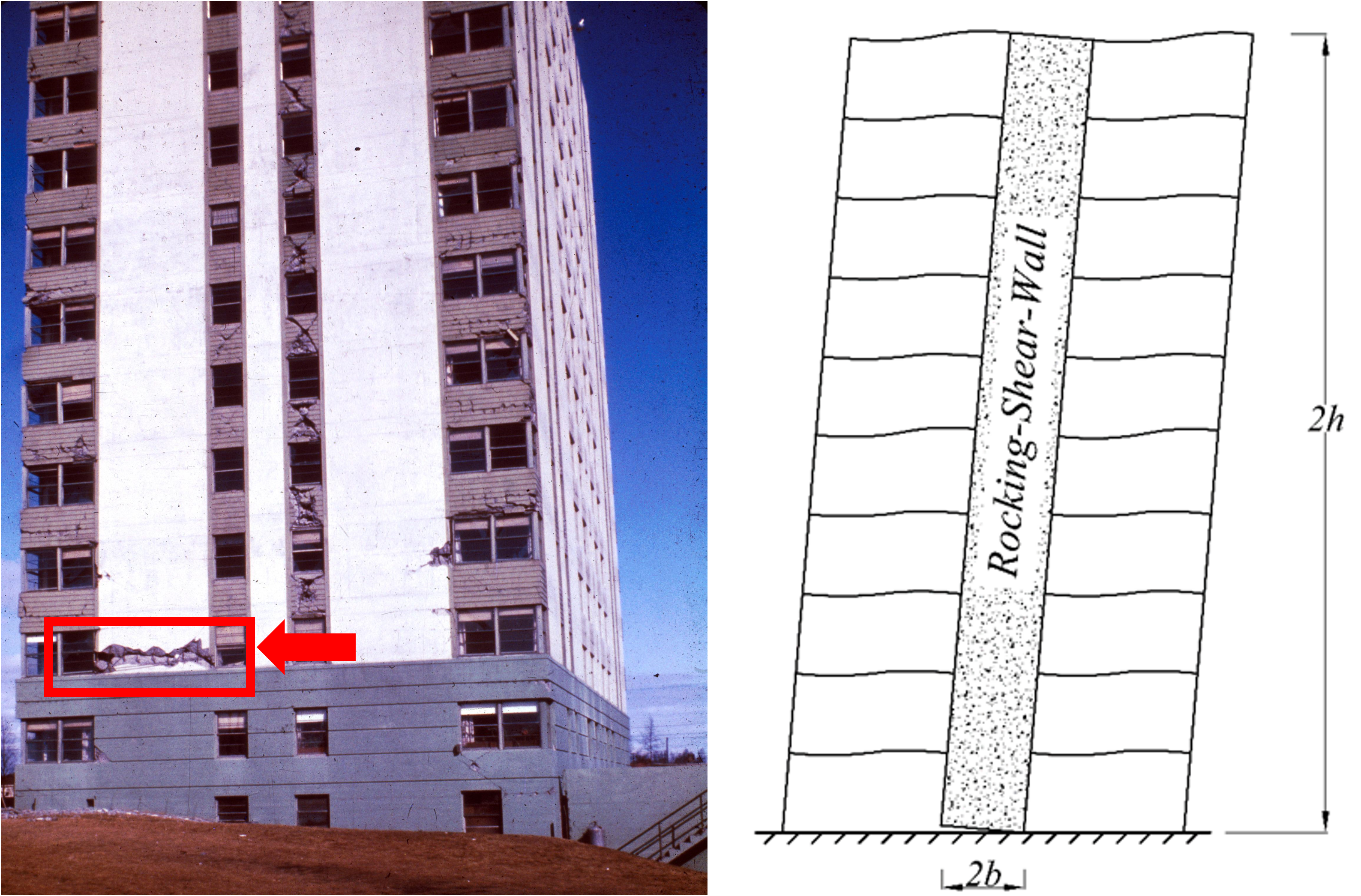}
\caption{Left: A fourteen-story reinforced concrete apartment building in Anchorage, Alaska, was severely damaged during the 1964 Alaska earthquake. One of the main exterior shear walls, shown in this figure, failed at the second floor, exposing the steel reinforced bars within the concrete (Image courtesy of the U.S. Geological Survey, \url{https://library.usgs.gov/} \cite{USGS}). Right: Schematic of the first mode deformation of a tall moment-resisting-frame with a rocking-shear-wall.}
\label{fig:1}
\end{figure}

To address significant seismic damage in moment-resisting frames, which sometimes led to weak-story failures, the idea of a rigid core system became increasingly popular. \cite{Paulay1969,Fintel1975,Emori1978,Bertero1980,Aktan1984}. Presence of large axial loads on the shear walls in a tall building reduces its ductility significantly; while the ductility demands are appreciable under long-duration pulse motions \cite{Paulay1986,Zhang2000}.  Moreover, the base of the core wall may suffer from cyclic degradation under prolonged shaking which usually results to permanent inelastic deformations. Such inelastic response may result to permanent drifts and lead to large repair costs; therefore, the entire design becomes unsustainable. An example of such failure after an earthquake is shown in Figure~\ref{fig:1} (left) for a fourteen-story moment-resisting frame with a fixed shear-wall building during 1964 Anchorage, Alaska earthquake \cite{NOAA}.

\begin{figure}[t]
\centering
\includegraphics[width=1.0\textwidth]{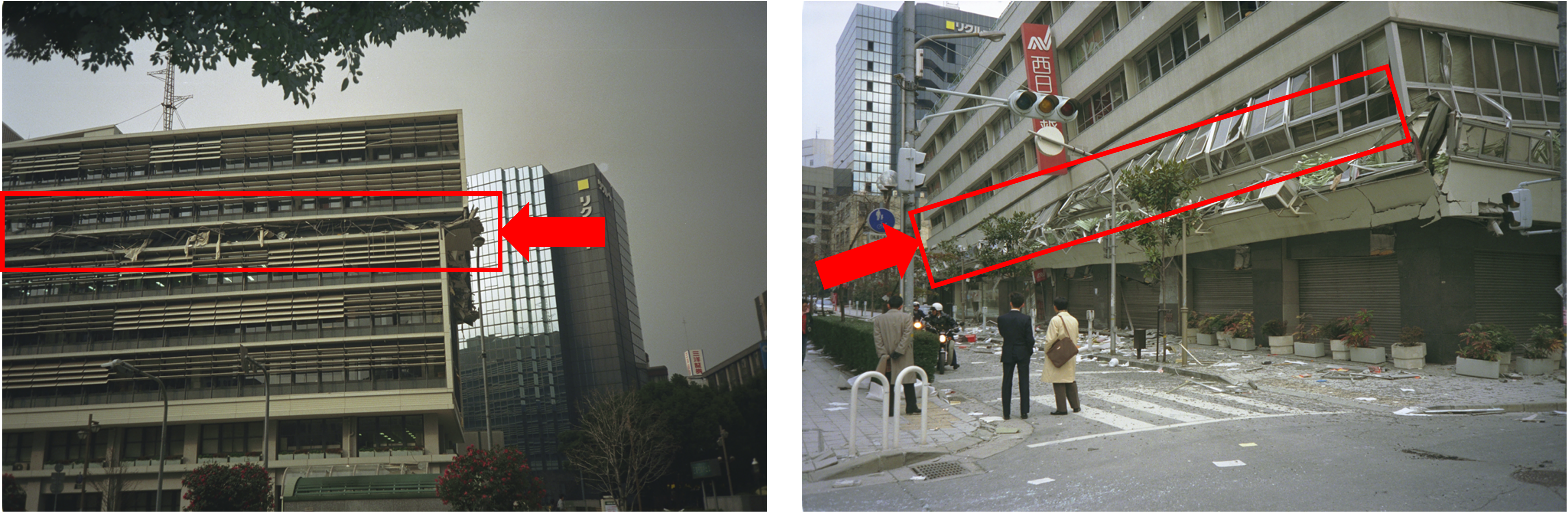}
\caption{Weak-story failure at the higher stories of the buildings after the 1995 Kobe, Japan Earthquake. (Image courtesy of the National Oceanic and Atmospheric Administration (NOAA) \cite{NOAA}).}
\label{fig:kobe}
\end{figure}

After 1994 Northridge, California earthquake followed by 1995 Kobe, Japan earthquake, coherent acceleration pulses (0.8-1.5 sec duration at that time) which result in large monotonic velocity, received revived attention. Makris (1996) \cite{makris1970near}, Alavi and Krawinkler (2004) \cite{alavi2004behavior} studied the destructive potential of pulse-like ground motions recorded near the causative fault of earthquakes.  In particular, several tall moment-resisting frames that had been designed in accordance with the existing seismic-code provisions exhibited a weak-story failure--in some cases several stories above the ground (see Figure \ref{fig:kobe}). 

Over the past few decades, there has been an increasing effort to highlight the unique advantages with allowing main vertical structural elements (such as piers in bridges or shear wall in buildings) to uplift and rotate above their foundation to mobilize a lower failure mechanism by design. The advantages of this approach is that the failures associated with cyclic degradation are essentially avoided; while, permanent displacements remain small due to the inherent recentering tendency of the rocking mechanism \cite{kelly1972mechanisms,beck1973seismic,RN62,RN7,RN86,ajrab2004rocking,RN5,Gioiella2018,Aghagholizadeh2022}.

Furthermore, with advances in construction methods and need of higher quality materials in complex engineering projects, interest in use of prefabricated structural elements is increasing. The advantages of this type of construction can be summarized as, higher quality structural members, decreasing the cost of (on-site) construction, and at the same time reducing construction time  \cite{lacey2018structural,liew2019steel,wang2023seismic}. Hence, use of vertical structural members that are not fixed at the their foundation have a great potential in prefabricated construction method.

Part of the reason for this articulated or semi-articulated seismic design alternative is that on several occasions, the further strengthening of the building with fixed-based shear walls leads to the attraction of larger seismic forces and the entire approach reaches an impasse given that the resulting forces that develop cannot be accommodated by cost-effective foundations \cite{Aghagholizadeh2022}. Another major issue that is a concern in multistory buildings in which their earthquake performance relies on ductile behavior is that after severe shaking the multi-story building may end up with appreciable permanent displacements and there is a need for re-centering which in most cases leads to demolition (an outcome against the emerging trends of functional recovery) as happened after the 2011 Christchurch, New Zealand earthquake \cite{elwood2013performance}.

This paper first provides a comprehensive literature review of rocking elements such as columns and shear walls. Then discusses different configurations of rocking shear walls. Lastly, the paper provides a numerical study for two steel moment-resisting frames (9 and 20 story) when they are coupled with a modular rocking shear wall under various seismic loading and compares their floor displacements and story drifts.

\section{Overview of the Literature}
\begin{figure}[b!]
\centering
\includegraphics[width=\textwidth]{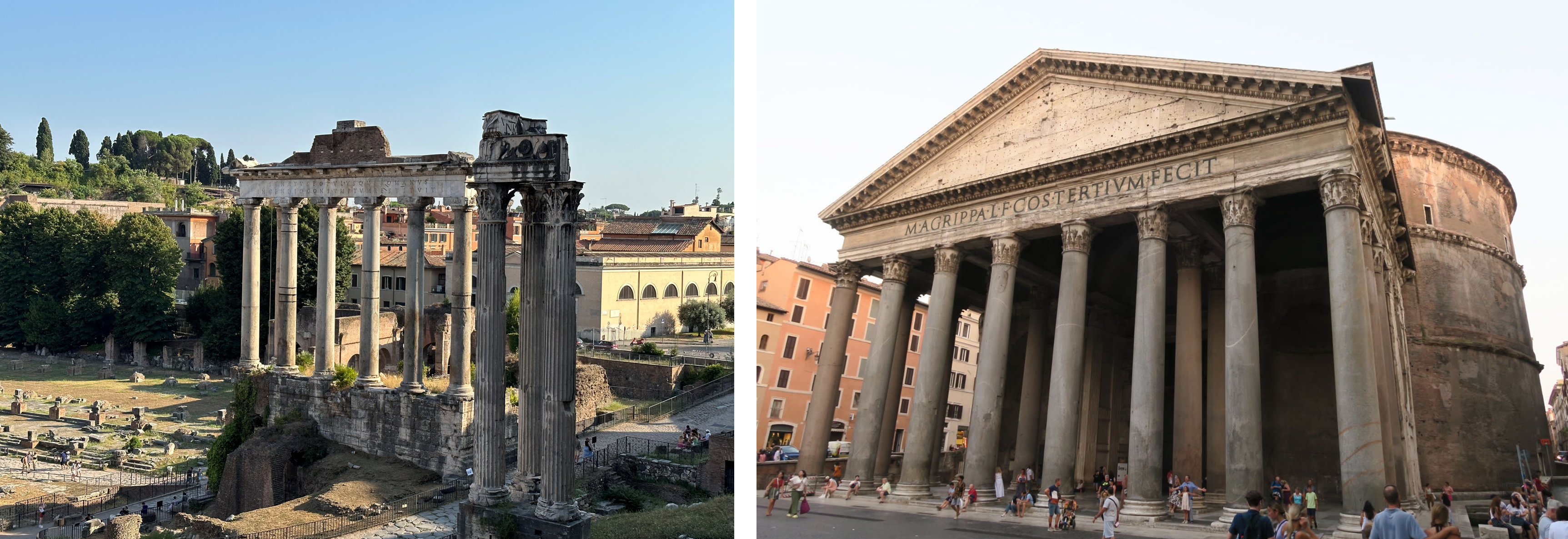}
\caption{Left: A view of columns in the Roman Forum, Rome, Italy. Right: The monolithic columns of Pantheon's portico in Rome, Italy.}
\label{fig:rome}
\end{figure}

Articulated construction of structural members dates back to ancient Greek and Roman buildings. Examples of such buildings are Acropolis of Athens, columns of remaining structures in the Roman Forum,  and columns of the Pantheon in Rome (shown in the Figure \ref{fig:rome}). The articulated construction of columns in these ancient buildings are either comprise of a monolithic column or multi-block columns \cite{stiros2020monumental}. What makes these ancient articulated construction interesting is that these structures survived several seismic motions during their life cycle. The seismic performance of these structures are caused by their unique properties, that is; re-centering is accomplished by their self-weight, negative stiffness nature of the rocking blocks, and the large size of the structures develops enough resistance to overturning mobilizing its rotational inertia \cite{makris2014half}.

In contemporary structural engineering, the concept of allowing a tall, slender structure to uplift and rock was first advanced and implemented in modern civil engineering in the late 1960s in New Zealand with the design and construction of the stepping piers of the South Rangitikei bridge \cite{beck1972seismic,beck1973seismic,bookKelly,skinner1993introduction}. This distinctive design was developed out of necessity because the piers of the South Rangitikei Bridge were over 75 meters tall, resulting in excessively large overturning moments at the pier foundations when using a conventional capacity design. The design of the South Rangitikei Bridge was developed by the Physics and Engineering Laboratory of Department of Scientific and Industrial Research (DSIR) in New Zealand \cite{beck1972seismic,beck1973seismic,bookKelly,skinner1993introduction,robinson1976extrusion}. Their efforts resulted to the development of the torsionally yielding steel dampers that was used to enhance the energy dissipation of the base of the stepping piers of the South Rangitikei Rail Bridge \cite{kelly1972mechanisms,skinner1974hysteretic,kelly1977earthquake}.

Another early investigation into the concept of shear walls capable of uplifting and rocking was conducted by Meek in 1978 \cite{meek1978dynamic}. Inspired by the seminal work on rocking blocks by Housner (1963) \cite{housner1963behavior} this study used a simplified analysis of the core rocking wall and a frame when the wall and footing rock on the soil. This study demonstrated that rocking walls significantly reduce the base shear and moment at the base of the wall compared to conventional fixed-base shear walls. Around the same time Clough and Huckelbridge explored the idea of steel frames that were free to uplift and rock above their foundation with shake table tests \cite{clough1977preliminary}. The test consists of a three-story single-bay steel frame with and without column uplift. The rocking steel frame of this study was pinned at the bottom, allowing free body rotation when the column has been uplifted. The experiment revealed that the presence of rocking reduced the seismic demand relative to a fixed base frame.

Despite the remarkable originality of these early works and their technical merit, these papers did not receive the attention it deserved, and it was some two decades later that the PRESSS (PREcast Seismic Structural Systems) Program \cite{RN62,priestley1996presss} reintroduced the concept of uplifting and rocking of the joint shear wall system \cite{Nakaki1999,priestley1999preliminary}.

Kurama et al. (1999, 2002) investigated behavior of unbonded post-tensioned precast concrete walls \cite{RN7,kurama2002seismic,kurama1999b}. These studies propose a design method based on an idealized trilinear relationship between base shear and roof drift. The trilinear relationship comprises four stages: the decompression state (when the wall begins to uplift), the softening state (the linear limit determined either by the gap opening of the walls or the nonlinear behavior of concrete under compression), the yielding state (where the strain in the post-tensioning steel first reaches the linear limit), and the failure state (where the wall ultimately fails). This study \cite{RN7} validates the analytical model using test results from the National Institute of Standards and Technology (NIST)  \cite{cheok1993model}.The verification of the analytical model compared with test results shows that the analytical model is reasonably agreed with the test in loading; however, the model is not accurate in unloading phase. Additionally, in Kurama et al. \cite{kurama2002seismic} the drift results of the analytical model, based on the equal displacement assumption, do not accurately predict the experimental results.

Holden et al. (2003) compared behavior of monolithic reinforced concrete walls with prestressed concrete walls using reversed cyclic quasi-static lateral loading \cite{holden2003seismic}. In this study two geometrically identical (half-scale) concrete wall were tested under quasi-static loading. One wall was conventional reinforced fixed-end shear wall, while the other wall was a partially prestressed precast wall which was free to uplift and rock on the pivoting points. For the rocking wall in this study assumed a bilinear elastic behavior which is inspired by the damage avoidance design (DAD) paradigm \cite{mander1997seismic} and it ignores effects of wall's mass inertia. Results from this study shows that partially prestressed rocking wall with energy dissipating devices achieved drift level on excess of 3\% with no visible damage. This study primarily focused on the advantages of self-centering walls compared to monolithic shear walls, particularly regarding wall damage after lateral loading and the comparison of residual displacements.

Ajrab et al. (2004) analyzed a rocking wall-reinforced concrete frame system with additional tendon system and dampers \cite{ajrab2004rocking}. To calculate the lateral capacity demand of the rocking wall-frame system, this study employed a capacity design approach. In the pre-rocking stage, the system behavior governed by structural flexibility, when uplifting of the wall initiated, based on the equilibrium of internal and external works, the base shear capacity of the system is calculated without considering effect of rotational inertia of the rocking wall. Then overall performance of the structure under MCE (Maximum Considered Earthquake) and MAE (Maximum Assumed Earthquake) is compared with the maximum displacement of the structure, calculated using time-history analysis. The results of designed structure under different ground motions showed that the adopted capacity-demand method predicts larger displacements in comparison to what was obtained from time-history analysis. Additionally, inter-story drifts are also reduced and became more uniformly distributed through the height of the building.

\begin{figure}[b!]
\centering
\includegraphics[width=1.0\textwidth]{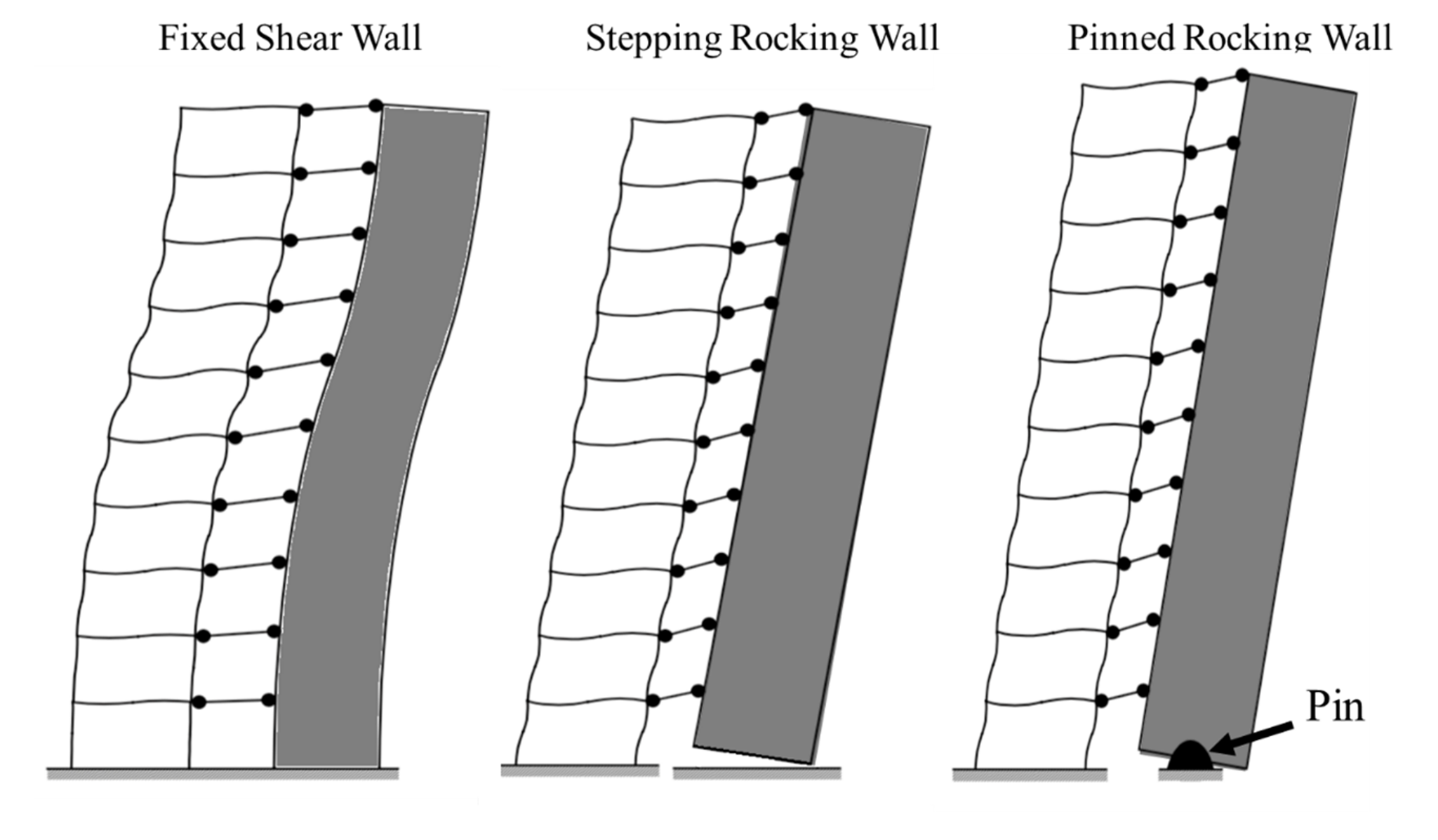}
\caption{Conventional Fixed shear wall (left) compared to stepping rocking wall (center) and Pinned rocking wall (right).}
\label{fig:wall_types}
\end{figure}

To strengthen moment-resisting frames to near-fault ground motion effects, Alavi and Krawinkler (2004) introduced pinned rocking wall system similar to the one shown in Figure \ref{fig:wall_types} (right) \cite{RN86}. Near-fault ground motions cause a highly non-uniform distribution of story ductility demand. Coupling the moment-resisting frame with the pinned wall in this study showed that this is an effective strategy that reduces drift demands of structures with a wide range of periods and various performance levels.

Resilience and serviceability of self-centering structural systems after major earthquakes were studied by Filiatrault et al. \cite{filiatrault2004development}. The paper points the main advantages of these systems as: their large lateral displacement capacity, the lack of structural damage associated with large displacements and their ability to return to the original position upon unloading \cite{filiatrault2004development}.

Lu (2005) studied behavior of rocking wall-frame system considering its 3D effect \cite{lu2005inelastic}. Purpose of the study was modeling wall's neural axis migration and showing its significance and assessment of 3D effect of the wall in order to control it. This study showed that the uncontrolled wall rocking can cause beam-wall connection failures. This study conducted an experiment on a six story high wall-frame system when it is excited with different ground motions. The main goal of the experimental work was to investigate the coupled system's maximum displacement. This paper concluded that most of the first-story lateral drift was attributed to the wall rigid body rotation about its pivoting points during inelastic response.

Restrepo and Rahman (2007) studied the effect of energy dissipators in self-centering walls \cite{restrepo2007seismic} in an experimental work. The study showed that even wall with no dampers will have no residual displacement after experiments.

Effect of post-tensioning a self-centering wall was investigated by Erkmen and Schultz (2009) \cite{erkmen2009self}. Results of the experiments showed that the rocking wall is capable of recentering, even if the post-tensioning force is died out after cycling loading.

Tozano et al. (2009) conducted an experimental work on a confined masonry wall that is free to rock \cite{toranzo2009shake}. To provide damping, the wall was equipped with energy dissipators. The mockup wall was build on 40\% scale. The structure showed a good performance under different levels of ground motions and met the design criteria. Results from this study showed that the rocking masonry wall was able to recenter after series of seismic loading and showed no visible damages even after maximum roof drift ratio of 2.\%.

Wada et al. (2011) \cite{wada2011seismic,RN5} retrofitted an existing eleven-story reinforced concrete moment resisting frame with a damped pinned-rocking wall. This building was located in the campus of Tokyo Institute of Technology in Japan. The analytical model of the building with wall is studied using ABAQUS software. In the building in order to take advantage of vertical displacement in the pinned-rocking-wall sets of vertical hysteretic dampers were used. The results showed that the retrofitted building experiences less drifts and the drifts are evenly distributed along the height of the building.

An experimental and numerical study on a 1:3 scaled structure with a rocking wall called propped rocking wall (PRW) conducted by Nicknam and Filiatrault (2014) \cite{nicknam2014numerical}. The experimental model's measured fundamental period was significantly larger than the numerical model. The displacement results were close but the numerical model damps out quicker than the experimental test. 

Wiebel and Christopoulos studied the controlled rocking steel braced frames (CRSBFs) for seismic resisting systems \cite{wiebe2015performance,wiebe2015performanceII}. In their studies they developed a framework to design these systems. In their study they showed how to design the base rocking joint to achieve a targeted response based on the results of series of analyses of single-degree-of-freedom systems with flag-shaped hystereses \cite{wiebe2015performance}. 

Grigorian and Grigorian recommended a design procedure called design-led analysis for the rocking-wall-moment-frame (RWMF) systems and studied few examples of generic structures \cite{grigorian2016performance,grigorian2016introduction}. The rocking-wall that was used in these analysis was a pinned-rocking wall. Their study was based mostly on capacity design which did not consider the rotational inertia of the wall under dynamic loading.

Nazari et al. (2017) investigated different precast rocking walls with various prestressing and tendon area configurations \cite{nazari2017single,nazari2019seismic,kalliontzis2022improving}. In terms of performance under different seismic loadings rocking walls performed satisfactorily and sustained negligible damages.

Gioiella et al. studied the implementation of rocking frames as a viable method in retrofitting of an existing building \cite{gioiella2017innovative,gioiella2018modal}. The retrofit of the existing building was conducted coupling the building with turss pinned rocking towers equipped with dampers. The study concluded that use of the proposed rocking system helped to reduce the horizontal displacements and accelerations.

Di Egidio et el. investigated seismic response of frames coupled with an external rocking wall and idea of adding inertes to the coupled system \cite{di2020seismic,di2021combined}. In their study they showed that rocking rigid block improves the seismic response of frame structure \cite{di2020seismic}, and it showed that use of inerters increases the effectiveness of coupling with rocking wall \cite{di2021combined}.

In series of publications Aghagholizadeh and Makris studied dynamics of a structural frame when it is coupled with rocking walls \cite{makris2017dynamics,aghagholizadeh2017seismic,makris2017earthquake,dissertation2018,makris2019effect,aghagholizadeh2021response,makris2017Bearthquake,Aghagholizadeh2022,agg2012relation,agg2016new,agg2018study,agg2018seismic}. In their studies they investigated a single-degree-of-freedom idealization of a moment-frame coupled with a rocking wall. They compared the results of different rocking wall configuration and confirmed the single-degree-of-freedom model against a multi-degree of freedom moment frame when it is coupled with a rocking wall. Some of the results presented in the next section of this paper is derived from the studies that has been conducted within these work \cite{makris2017dynamics,aghagholizadeh2017seismic,makris2017earthquake,dissertation2018,makris2019effect,aghagholizadeh2021response,makris2017Bearthquake,Aghagholizadeh2022}.

Most of these aforementioned studies introduce the unique advantages of rocking action by referencing the seminal paper by Housner \cite{housner1963behavior}, who noticed that tall, slender, free-standing columns, while they can easily uplift even when subjected to a moderate ground acceleration (uplifting initiate when $\Ddot{u}_g>g\times(base/height)$); they exhibit remarkable seismic stability due to a size-frequency scale effect. In his 1963 paper Housner shows that there is a safety margin between uplifting and overturning and that as the size of the free-standing column increases or the frequency of the excitation pulse increases, this safety margin increases appreciably to the extent that large free-standing columns enjoy ample seismic stability. Makris \cite{makris2014half,makris2014role} explained that as the size of the free-standing rocking column increases, the enhanced seismic stability primarily originates from the difficulty to mobilize the rotational inertia of the column (wall) which increases with the square of the column (wall) size. Further studies by Makris and Vassiliou \cite{makris2014some,vassiliou2015dynamics} showed that as the size of the column (wall) increases, the resistance to mobilize the rotational inertia increases to such an extent, that the effect of vertical tendons becomes increasingly marginal.

\section{Comparison of Rocking Shear Walls for Different Configurations}
\label{section:comp}
\subsection{Stepping Rocking Wall and Pinned Rocking Wall}
Two of the major configurations for rocking walls are pinned rocking wall (shown in Figure \ref{fig:wall_types} right), and stepping rocking wall (shown in Figure \ref{fig:wall_types} center). Depending of the objective both have advantages and dis-advantages. Implementing the pinned rocking wall helps to avoid one of the weak points of a stepping rocking wall which is damage in its pivoting points caused by impact to the foundation. But having a stepping rocking wall provides some energy dissipation when the wall changes its pivoting corner from one side to the other which is caused by the impact.

With the assumption that the rocking wall behaves similar to a rigid wall (since the wall is not fixed in the bottom and for low to mid-rise buildings, this is a reasonable assumption) and the wall-spring connecting arm is long enough, the system shown in Figure \ref{fig:sdofs} is a single-degree-of-freedom system where the lateral translation of the mass, $u$,
is related to the rotation of the stepping rocking wall, $\theta$ \cite{aghagholizadeh2017seismic}.

For bilinear force-displacement behavior of the nonlinear spring Bouc-Wen model is selected \cite{bouc1967forced,wen1976method}.  Force that develops in the nonlinear spring can be shown as:

\begin{figure}[b!]
\centering
\includegraphics[width=1.0\textwidth]{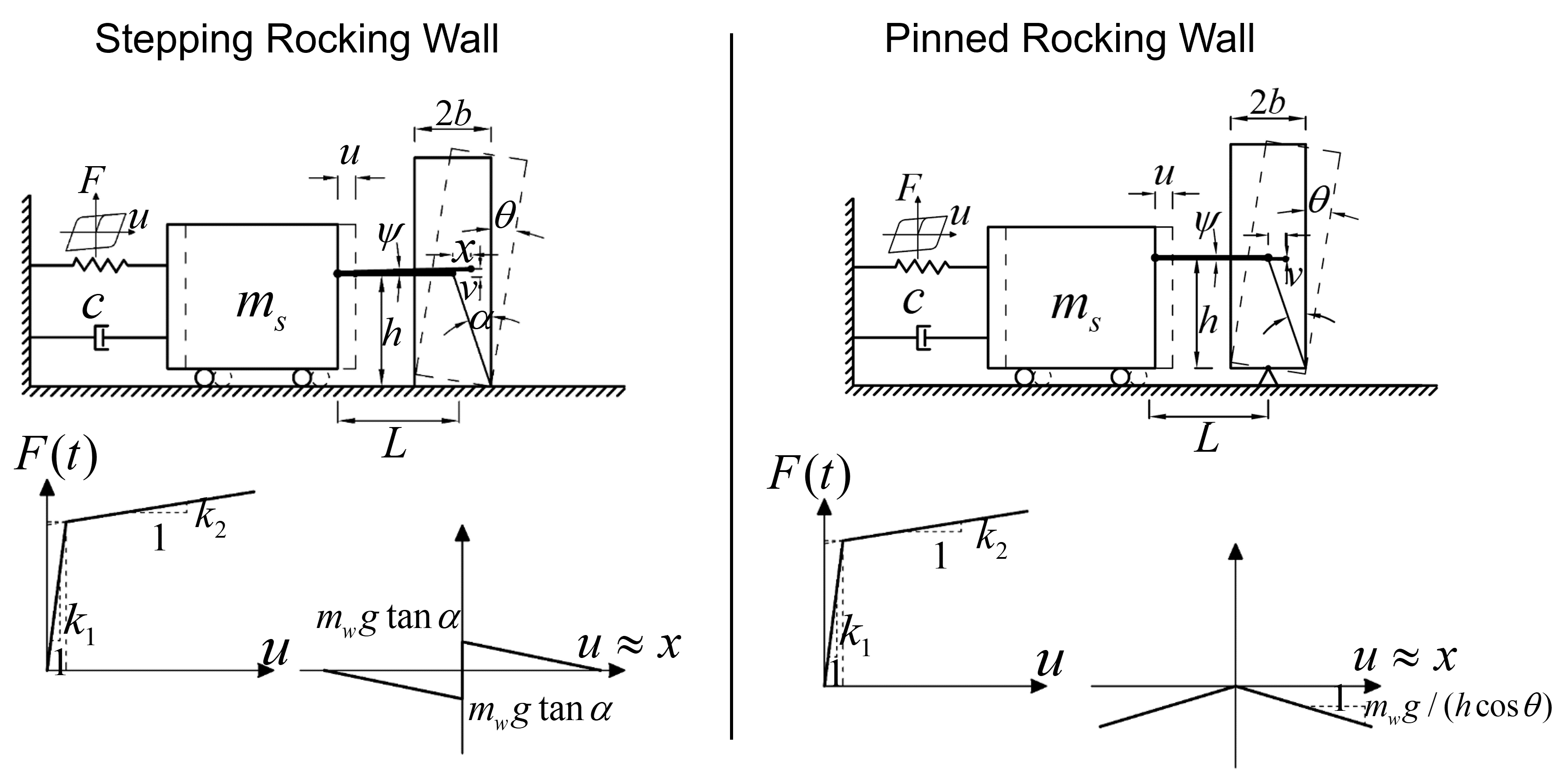}
\caption{Top row: Single-degree-of-freedom idealization of a yielding frame when it is coupled to a stepping rocking wall (left) and pinned rocking wall (right). Bottom row: Bilinear idealization with its control parameters for the nonlinear spring model and force-displacement diagram of the stepping rocking wall (left) and pinned rocking wall (right).}
\label{fig:sdofs}
\end{figure}

\begin{equation} \label{eq:f_s}
    F_s(t) = ak_1u(t) + (1-a)k_1u_yz(t),
\end{equation}
where $a=k_2/k_1$ is the post–to-pre yielding stiffness ratio and $-1\leqslant z(t)\leqslant 1$ is a dimensionless internal variable described by:
\begin{equation} \label{eq:z}
\dot{z}(t)=\dfrac{1}{u_y}[\dot{u}(t)-\beta \dot{u}(t) |z(t)|^n-\gamma|\dot{u}| z(t) |z(t)|^{n-1} ]
\end{equation}
In equation (\ref{eq:z}), constants $\beta$, $\gamma$ and $n$ are model parameters that dictates the shape of the hysteretic loop \cite{bouc1967forced,wen1976method,baber1981random}.

Using equilibrium of forces it can be shown the equation of motion for positive and negative rotation of the stepping rocking wall as follows (for detailed calculation of the equations of motion see \cite{aghagholizadeh2017seismic,2017eqtongji}):
\begin{equation}  \label{eq:12}
\begin{aligned}
\theta\geq0\\
&\Big(\dfrac{4}{3}+\sigma\cos^2(\alpha-\theta)
\Big)\ddot{\theta} \\
&
+\sigma \cos(\alpha-\theta) \Big[a\omega_1^2\big(\sin\alpha-sin(\alpha-\theta)\big)+2\xi\omega_1\dot{\theta}\cos(\alpha-\theta)
+\dot{\theta}^2 \sin(\alpha-\theta)
\\
&+(1-a)\omega_1^2\dfrac{u_y}{R}z(t)
\Big]={}-\dfrac{g}{R}\Big[(\sigma+1)\dfrac{\ddot{u}_g}{g}\cos(\alpha-\theta)+\sin(\alpha-\theta)\Big],
\end{aligned}
\end{equation}
and, 
\begin{equation}  \label{eq:13}
\begin{aligned}
\theta<0\\
&\Big(\dfrac{4}{3}+\sigma\cos^2(\alpha+\theta)
\Big)\ddot{\theta} \\
&
-\sigma \cos(\alpha+\theta) \Big[a\omega_1^2\big(\sin\alpha-sin(\alpha+\theta)\big)-2\xi\omega_1\dot{\theta}\cos(\alpha+\theta)
+\dot{\theta}^2 \sin(\alpha+\theta)
\\
&-(1-a)\omega_1^2\dfrac{u_y}{R}z(t)
\Big]={}\dfrac{g}{R}\Big[-(\sigma+1)\dfrac{\ddot{u}_g}{g}\cos(\alpha+\theta)+\sin(\alpha+\theta)\Big],
\end{aligned}
\end{equation}

\noindent where wall of size $R=\sqrt{b^2+h^2}$, slenderness, $\tan\alpha=b/h$, $\sigma=\frac{spring~mass}{wall~mass}= m_s/m_w$ is the mass ratio parameter, $\omega_1=\sqrt{k_1/m_s}=$ the pre-yielding undamped frequency and $\xi=\dfrac{c}{2m_s \omega_1}=$ the pre-yielding viscous damping ratio of the SDOF oscillator. Equations (\ref{eq:12}) and (\ref{eq:13}) are the equation of motion for positive and negative rotations of the coupled system shown in Figure \ref{fig:sdofs} (left).

Similarly equation of motion for a pinned rocking wall when it is coupled to a nonlinear spring can be shown as follows (for detailed calculation of the equations of motion see \cite{aghagholizadeh2017seismic}):
\begin{equation} \label{eq:24}
\begin{split}
&\big[ \dfrac{1}{3}+(1+\sigma\cos^2\theta)\cos^2\alpha \big]\ddot{\theta}
\\
&+\sigma \cos^2\alpha\cos\theta \Big[
(a \omega^2_1-\dot{\theta}^2)\sin\theta+
(1-a)\omega^2_1 \dfrac{u_y}{R\cos\alpha}z(t)+
2\xi\omega_1\dot{\theta}\cos\theta  \Big]
\\
&=-\dfrac{g}{R}\cos\alpha \Big[ (\sigma+1)\dfrac{\ddot{u}_g}{g}\cos\theta-\sin\theta \Big],
\end{split}
\end{equation}
where $\omega_1=\sqrt{k_1/m_s}$ = the pre-yielding undamped frequency and $\xi=\dfrac{c}{2m_s \omega_1}$ = the viscous damping ratio of the SDOF oscillator (as in the previous case). Equation (\ref{eq:24}) is the equation of motion for both positive and negative rotations of the coupled system shown in Fig.~\ref{fig:sdofs} (right).

Comparing equations of motion for stepping (\ref{eq:12} and \ref{eq:13}) with with pinned wall (\ref{eq:24}) it can be observed that the self-weight of the pinned rocking wall works against the wall's stability (term, $\frac{g}{R}\sin{\theta}$). This is one of the benefits of employing stepping rocking wall to pinned rocking wall, this way the wall have more tendency to recenter to its origin only employing its self-weight \cite{makris2017dynamics}. Hence, the stepping rocking wall will be more effective in reducing the residual displacement of the structure \cite{aghagholizadeh2017seismic}).

\subsection{Restrained Rocking Wall}
Another configuration of rocking wall is adding prestressing strands to the rocking wall and anchoring it to the foundation \cite{makris2017earthquake}. Some of the main advantages of using tendons to anchor the wall is that this configuration prevents possible sliding of the wall on top of its foundation and increases the vertical stiffness of the rocking-wall.

\begin{figure}[t!]
\centering
\includegraphics[scale=.75]{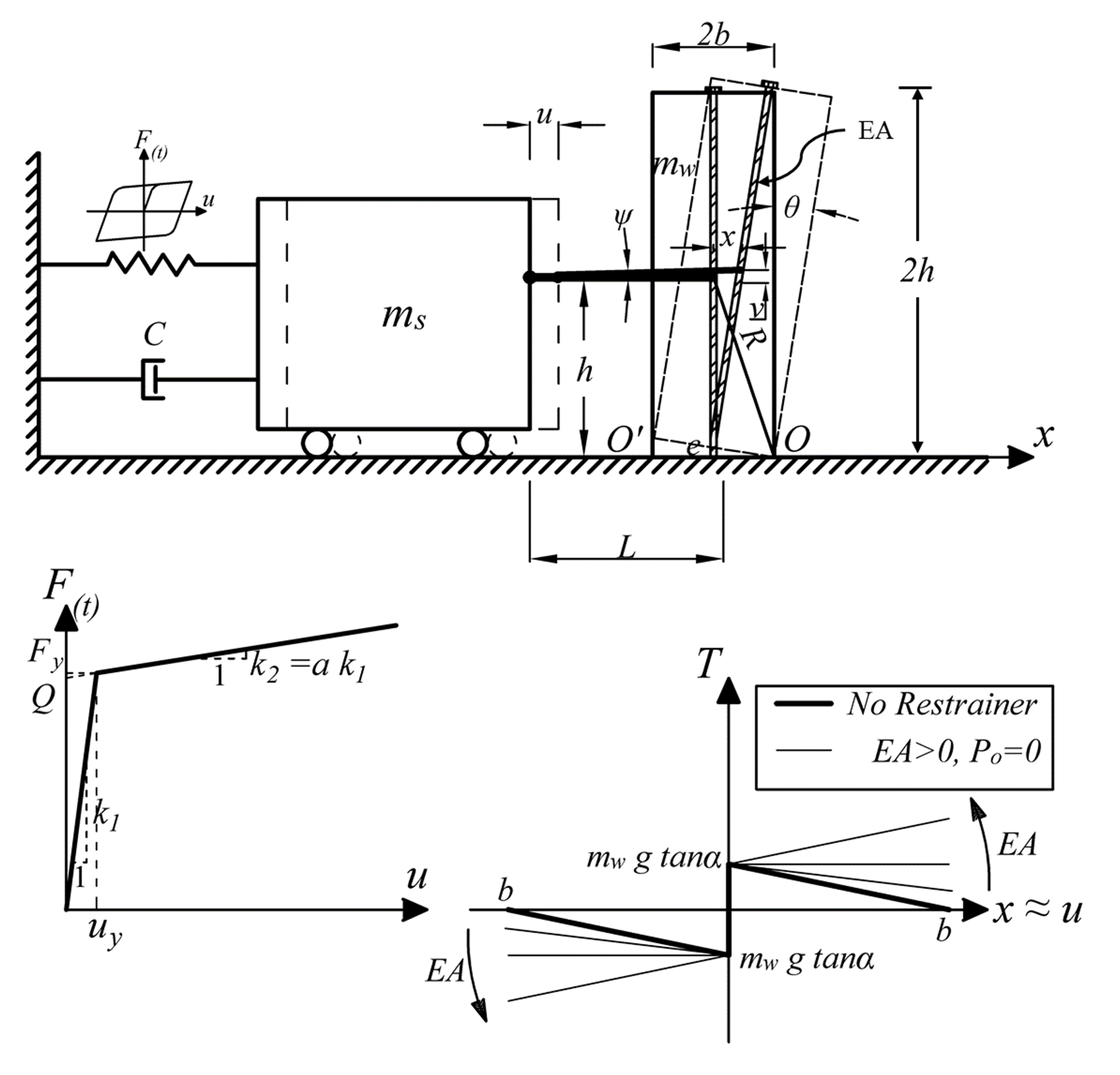}
\caption{Yielding single‐degree‐of‐freedom oscillator coupled with a vertically restrained stepping rocking wall.}
\label{fig:sdof_restrained}
\end{figure}

Figure (\ref{fig:sdof_restrained}) top presents an idealized single-degree-of-freedom system of a nonlinear spring and mass system representing a moment-resisting frame coupled with a vertically restrained stepping rocking wall. Bottom row of Figure (\ref{fig:sdof_restrained}) on the left shows bi-linear force-displacement relation assigned to the spring, and on the right side force-displacement diagram of the restrained stepping rocking wall.

As it shown in the bottom right of Figure (\ref{fig:sdof_restrained}), with the increase in the elasticity (EA) of the restrainer, the lateral stiffness of the vertically restrained stepping rocking shear wall gradually shifts from negative to positive \cite{vassiliou2015dynamics,makris2015dynamics}. Hence, it should be noted that if using a shear wall with negative stiffness is one of the objectives in the design, one have to limit amount of the tendons uses in the stepping rocking wall.

Using equilibrium of forces the equation of motion for positive and negative rotation of the vertically restrained stepping rocking wall show as follows (for detailed calculation of the equations of motion see \cite{makris2017earthquake}):

\begin{equation}  \label{eq:rest_pos}
\begin{aligned}
\theta\geq0\\
&\Big(\dfrac{4}{3}+\sigma\cos^2(\alpha-\theta)
\Big)\ddot{\theta}
+\sigma \cos(\alpha-\theta) \Big[a\omega_1^2\big(\sin\alpha-sin(\alpha-\theta)\big)+2\xi\omega_1\dot{\theta}\cos(\alpha-\theta)\\
&+\dot{\theta}^2 \sin(\alpha-\theta)
+(1-a)\omega_1^2\dfrac{u_y}{R}z(t)
\Big]
\\
&={}-\dfrac{g}{R}\Big[(\sigma+1)\dfrac{\ddot{u}_g}{g}\cos(\alpha-\theta)+\sin(\alpha-\theta)
+\sin\alpha \sin\theta (\frac{1}{2}\frac{EA}{m_wg} \tan\alpha+\frac{P_o}{m_wg}\frac{1}{\sqrt{2}\sqrt{1-\cos\theta}})\Big],
\end{aligned}
\end{equation}
and, 
\begin{equation}  \label{eq:rest_neg}
\begin{aligned}
\theta<0\\
&\Big(\dfrac{4}{3}+\sigma\cos^2(\alpha+\theta)
\Big)\ddot{\theta}
-\sigma \cos(\alpha+\theta) \Big[a\omega_1^2\big(\sin\alpha-sin(\alpha+\theta)\big)-2\xi\omega_1\dot{\theta}\cos(\alpha+\theta)\\
&+\dot{\theta}^2 \sin(\alpha+\theta)
-(1-a)\omega_1^2\dfrac{u_y}{R}z(t)
\Big]
\\
&={}\dfrac{g}{R}\Big[-(\sigma+1)\dfrac{\ddot{u}_g}{g}\cos(\alpha+\theta)+\sin(\alpha+\theta)
-\sin\alpha \sin\theta (\frac{1}{2}\frac{EA}{m_wg} \tan\alpha+\frac{P_o}{m_wg}\frac{1}{\sqrt{2}\sqrt{1-\cos\theta}}) \Big], 
\end{aligned}
\end{equation}
in which $EA$ is the axial stiffness of the vertical strand that can be prestressed with a prestressing force $P_o$.

\begin{figure}[t!]
\centering
\includegraphics[scale=.6]{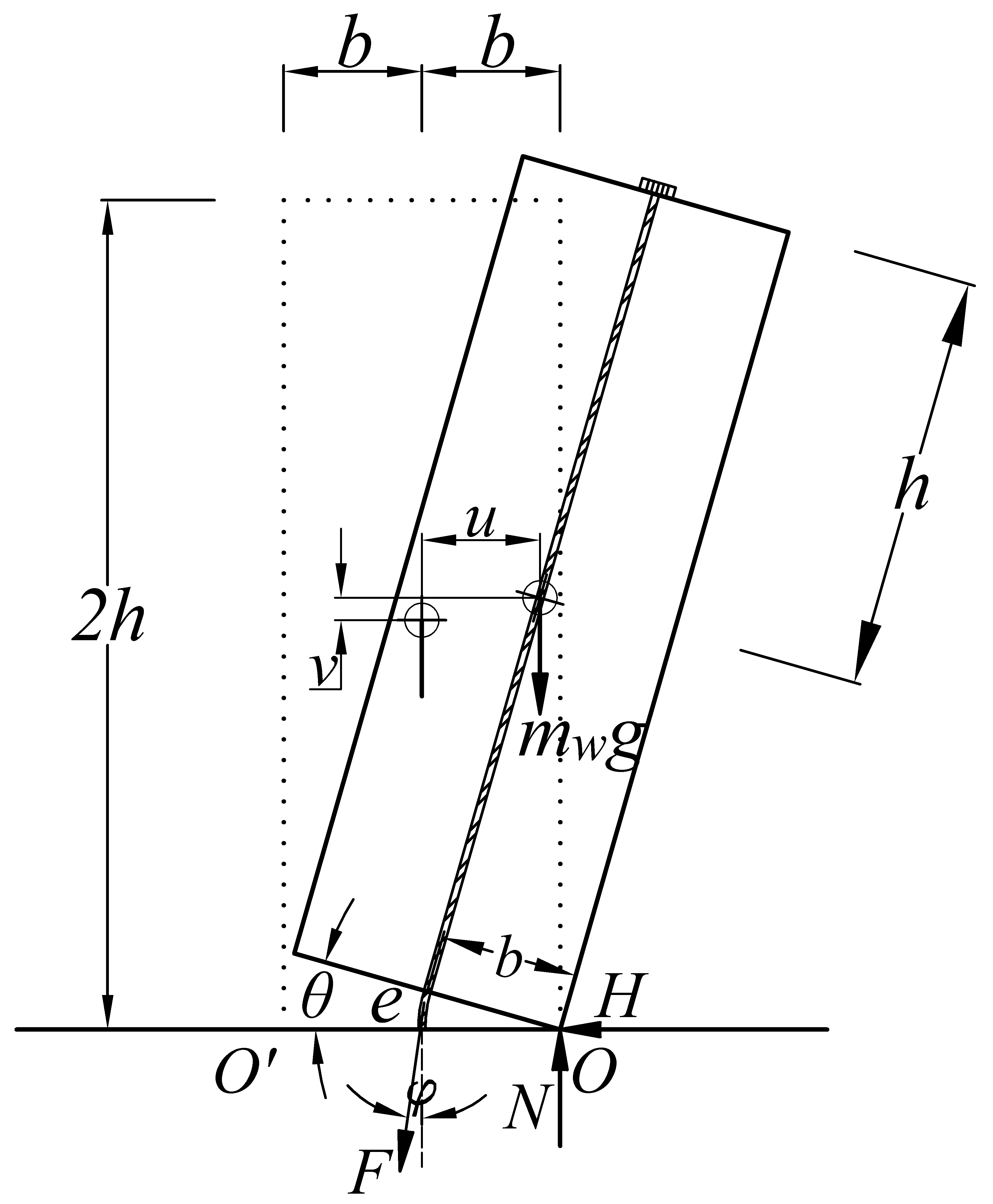}
\caption{Free-body diagram of a rocking wall with an elastic tendon passing through its center-line (without considering the spring connection).}
\label{fig:sdof_restrained_detail}
\end{figure}

Minimum ground acceleration $\ddot{u}_{g~min}$ that is required to initiate a vertically restrained shear wall is increased proportionally with the magnitude of the prestressing force $P_o$ \cite{makris2017earthquake}:
\begin{equation}  \label{eq:min_acc}
\ddot{u}_{g~min} = \dfrac{g~tan\alpha}{\sigma+1}\big(1+\dfrac{P_o}{m_w~g}
\big).
\end{equation}

Using equilibrium of forces at the pivoting corner of a rocking shear wall, shown in Figure (\ref{fig:sdof_restrained_detail}), the normal force, $N$, at the pivoting corner of the shear wall normalized with respect to the wall mass $m_w$, can be calculated as follows \cite{makris2017earthquake},
\begin{equation}  \label{eq:norml_force}
\begin{aligned}
\theta\geq0\\
&\frac{N(t)}{m_wg}=1+\frac{R}{g} \big[\ddot{\theta} \sin(\alpha-\theta)+ \dot{\theta}^2 \cos(\alpha-\theta)  \big]+\frac{1}{2}\frac{EA}{m_wg}\tan\alpha \sin\theta+\frac{1}{\sqrt{2}}\frac{P_o}{m_wg}\sqrt{1+\cos\theta}.
\end{aligned}
\end{equation}

\begin{figure}[t!]
\centering
\includegraphics[width=\textwidth]{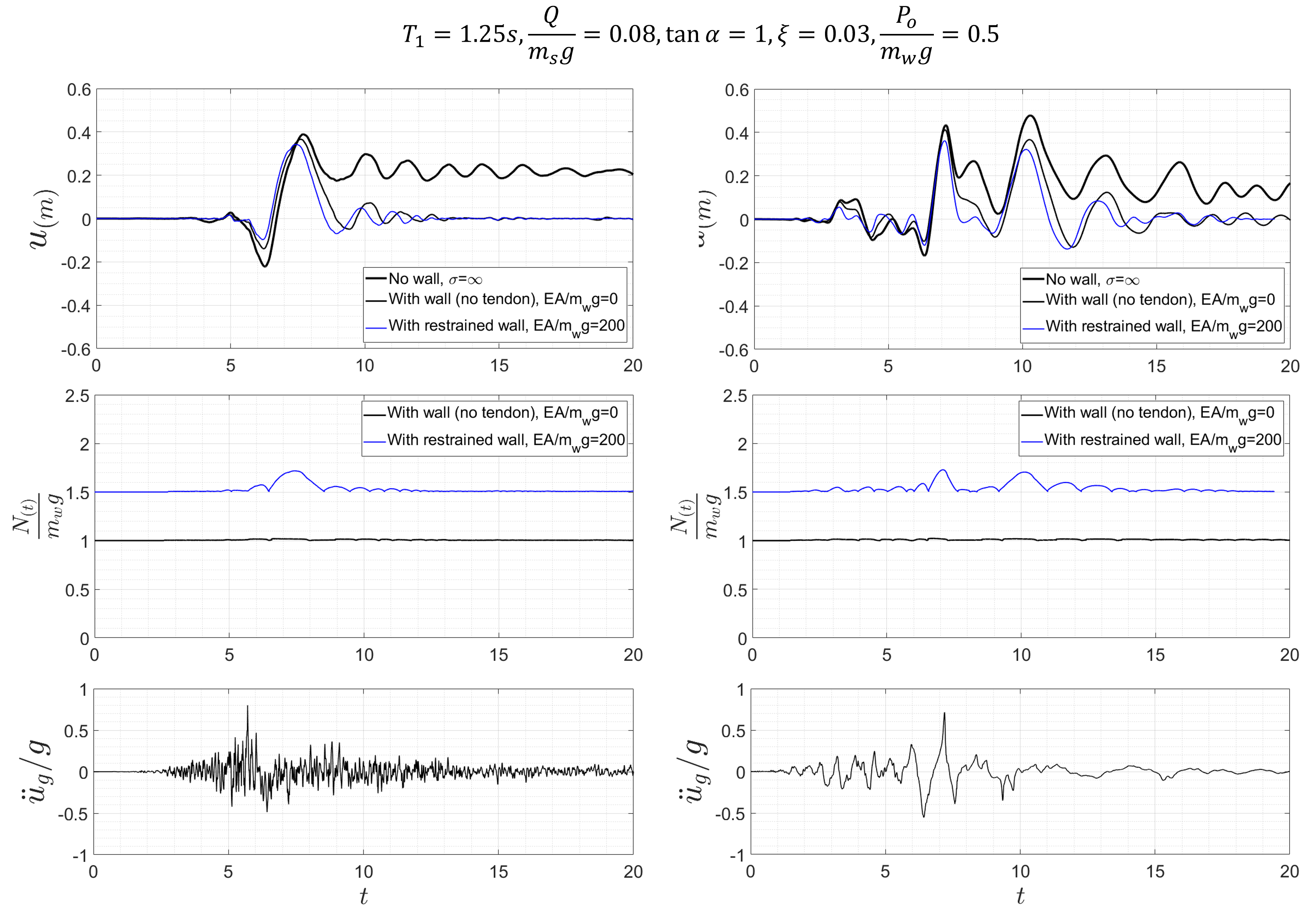}
\caption{Time-history analysis of a nonlinear SDOF oscillator coupled with a vertically restrained stepping rocking wall with
preyielding period, T1 = 1.25 seconds, normalized strength $Q/ms$ = 0.08g, wall size ratio, $\omega_1/p$ = 10, and structure-to-wall mass ration, $\sigma$ = 5. When subjected to the North-South component of NAR station ground motion recorded during 2023 Kahramanmaras, Turkiye (left) and REHS ground motion recorded during the 2011 Christchurch, New Zealand (right) earthquakes.}
\label{fig:response}
\end{figure}

Responses, $u(t)$, and normal forces at the pivoting point,$N(t)$, of a SDOF system when it is coupled with a rocking wall for the cases of with and without vertical tendon is shown in Figure (\ref{fig:response}). The plots on the left are responses when the system is subjected to the North-South component of NAR station ground motion recorded during 2023 Kahramanmaras, Turkiye earthquake and in the right side when the coupled system is subjected to the REHS ground motion recorded during the 2011 Christchurch, New Zealand earthquake. The solid heavy black line in the figure represents response of a nonlinear spring, light black line is the response of the SDOF system when it is coupled to a stepping rocking wall, and the blue line shows the response of the SDOF spring when it is coupled to a vertically restrained stepping rocking wall with pretensioned ($\frac{P_o}{m_wg} = $0.5) stiff tendons ($\frac{EA}{m_wg} = $200). The results shows that the use of a stepping rocking wall helps the SDOF to recenter to its origin, hence there is no residual displacement. The maximum displacement of the systems with rocking walls for both cases are smaller when it is compared to SDOF spring. 

Results in Figure (\ref{fig:response}) shows that by increasing the axial stiffness, $EA$, of the vertical tendon one increases the lateral stiffness of the entire structural system; at the same time, vertical tendons increase the vertical reactions at the pivoting corners (at some cases by more than 50\% \cite{makris2017earthquake}). This increases the possibility of additional damage to the pivoting corners of the shear walls which leads to an unsustainable design. Hence, when prestressing tendons are used, it is important to further investigate the effect of additional forces in the pivoting corners and retrofit the pivoting points to prevent generation of excessive damages.

\section{Numerical Analysis of Multi-Story Moment Resisting Frames When Coupled with Modular Stepping Rocking Walls}
In this section, seismic response of moment-resisting-frames (MRF) coupled with modular rocking shear walls, using OpenSees \cite{opensees} is investigated. 

\begin{figure}[t!]
\centering
\includegraphics[width=\textwidth]{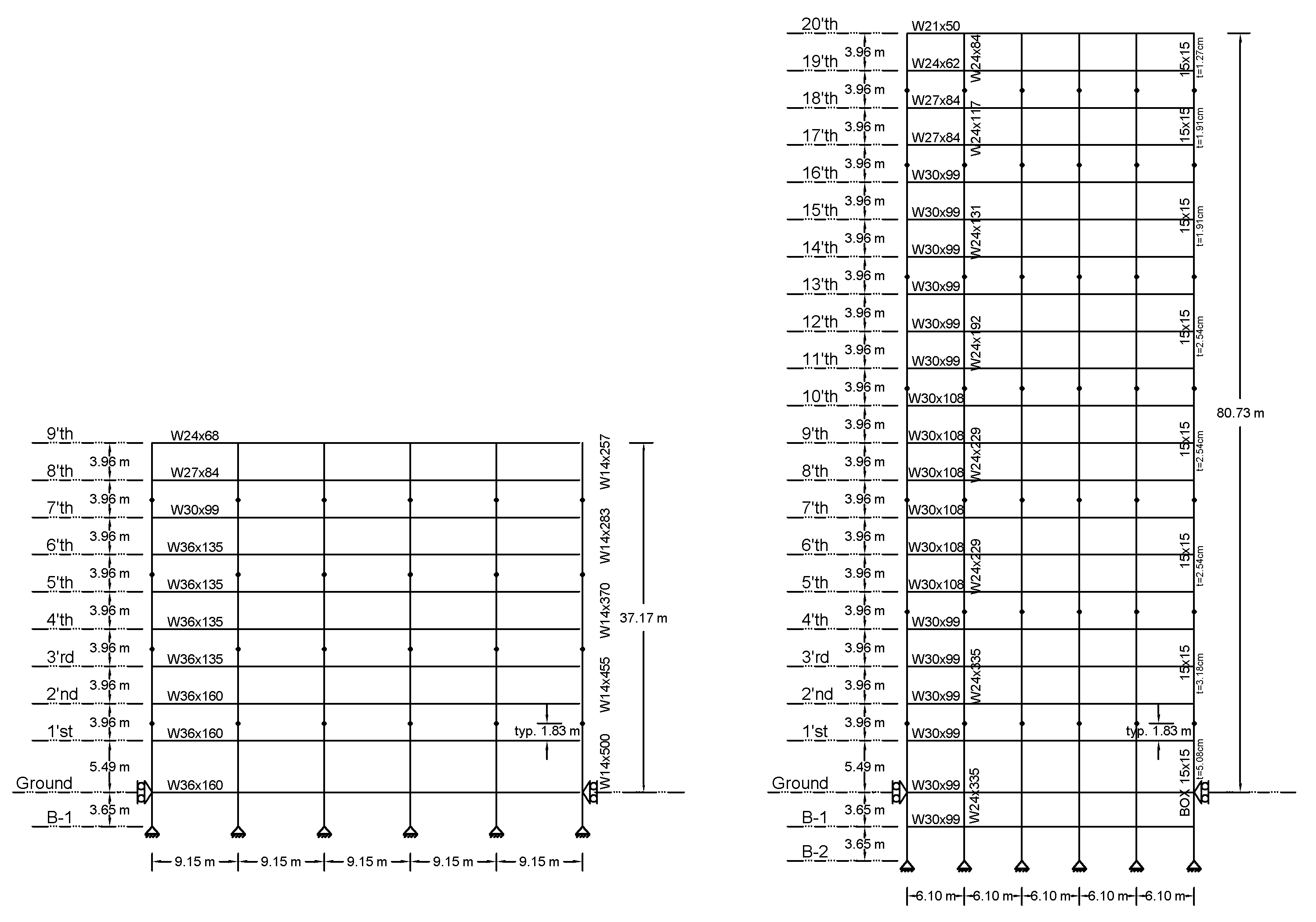}
\caption{Nine‐story (left) and twenty-story (right) moment‐resisting steel frame designed for the SAC Phase II Project.}
\label{fig:mrfs}
\end{figure}

The structures that are chosen for this study are moment resisting steel frames designed for the SAC Phase II Project \cite{SAC}. These buildings are well-known benchmark structures to the literature \cite{gupta1999seismic,gupta2000behavior,ohtori2004benchmark,makris2017earthquake,Aghagholizadeh2022}. The structures designed to meet the seismic code (pre‐Northridge Earthquake) and represents typical medium‐rise buildings designed for the greater area of Los Angeles, California. MRFs and their cross sectional information are shown in Figure (\ref{fig:mrfs}). The first three natural period of vibrations for the 9-story building is $2.26$ s, $0.874$ s, and $0.488$ s, and for the 20-story building is $3.83$ s, $1.33$ s, and $0.769$ s.

\begin{figure}[t!]
\centering
\includegraphics[width=\textwidth]{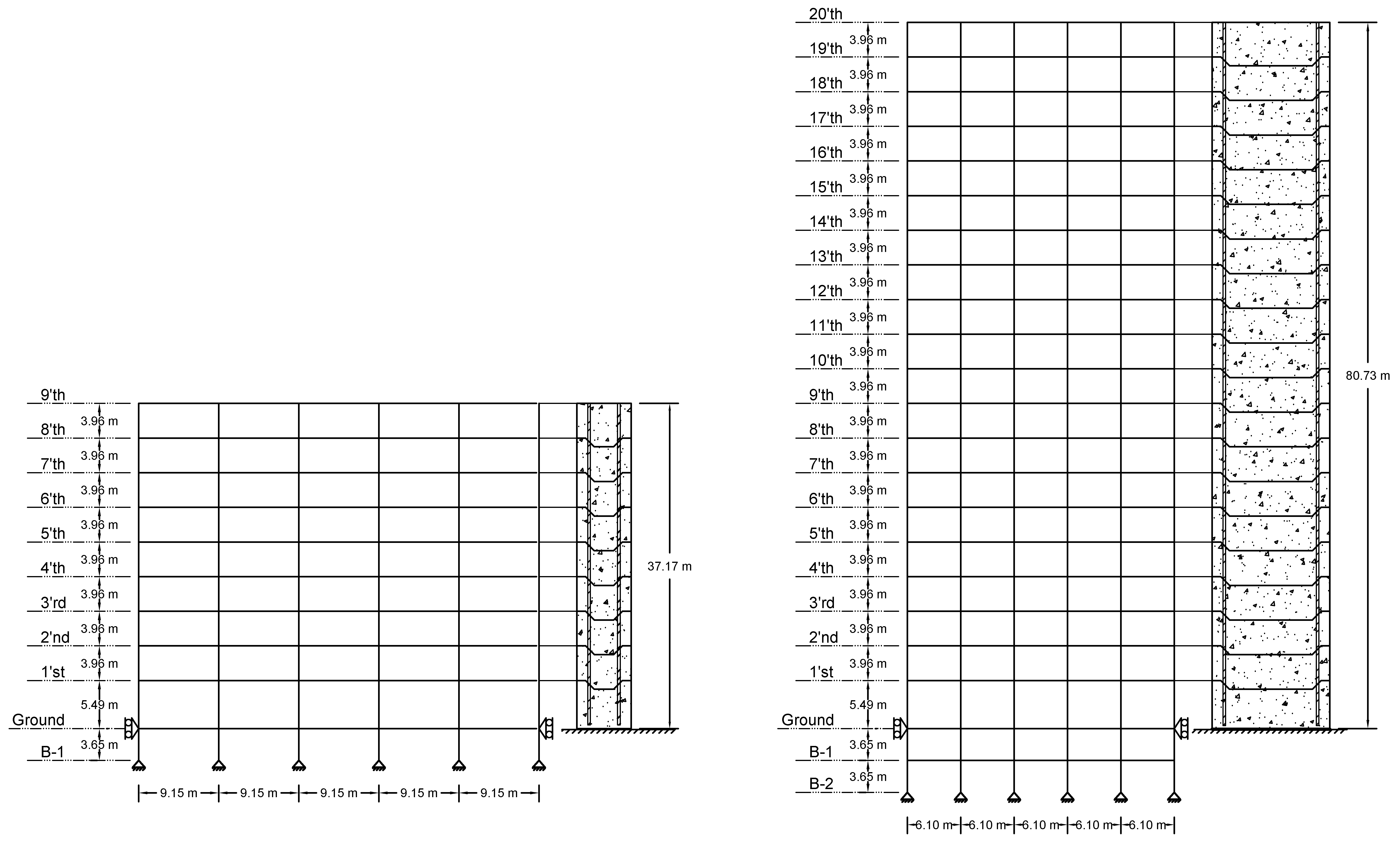}
\caption{Nine‐story (left) and twenty-story (right) moment‐resisting steel frames coupled with modular stepping rocking shear wall.}
\label{fig:segmental_wall}
\end{figure}

Considering the characteristics of rocking shear walls that discussed in section (\ref{section:comp}) of this paper, the nine-story and twenty-story frames are retrofitted with a modular stepping rocking shear wall using axial link elements between shear wall and the frame. The proposed system (shown in Figure \ref{fig:segmental_wall}) is comprised of wall segments that are puzzle shaped to provide mechanical connection between each wall parts. In addition to the mechanical connections, vertical prestressing strands are used to tie the wall segments together. These tendons pass through all walls and anchored inside the base wall above the foundation. This will increases the lateral stiffness of the wall that helps the wall act as a rigid block (which can be tuned with amount of prestressing applied) without adding extra stress at the pivoting corners.  

In order to model a rocking shear wall in OpenSees, a zero-length fiber section is define at the base of the wall between the wall and the ground surface. The cross-section of the zero-length rocking surface is defined, using of a fiber section with nonlinear, elastic-no-tension (ENT) material that is available in Opensees \cite{vas2017finite,aghagholizadeh2020finite}.

In order to consider energy dissipation of the rocking motion during the impact, similar to \cite{vas2017finite,aghagholizadeh2020finite,emi2024}, dissipative time-stepping integration procedure of Hilber-Hughes-Taylor \cite{HHT1977} (HHT) is being used. HHT damping effect is function of dissipation factor $a'_d$ and the time of the integration. For this study the HHT time step is selected as~$10^{-4}$ s and dissipation factor $a'_d$ is selected as~$-1/3$.

\begin{figure}[t!]
\centering
\includegraphics[width=0.78\textwidth]{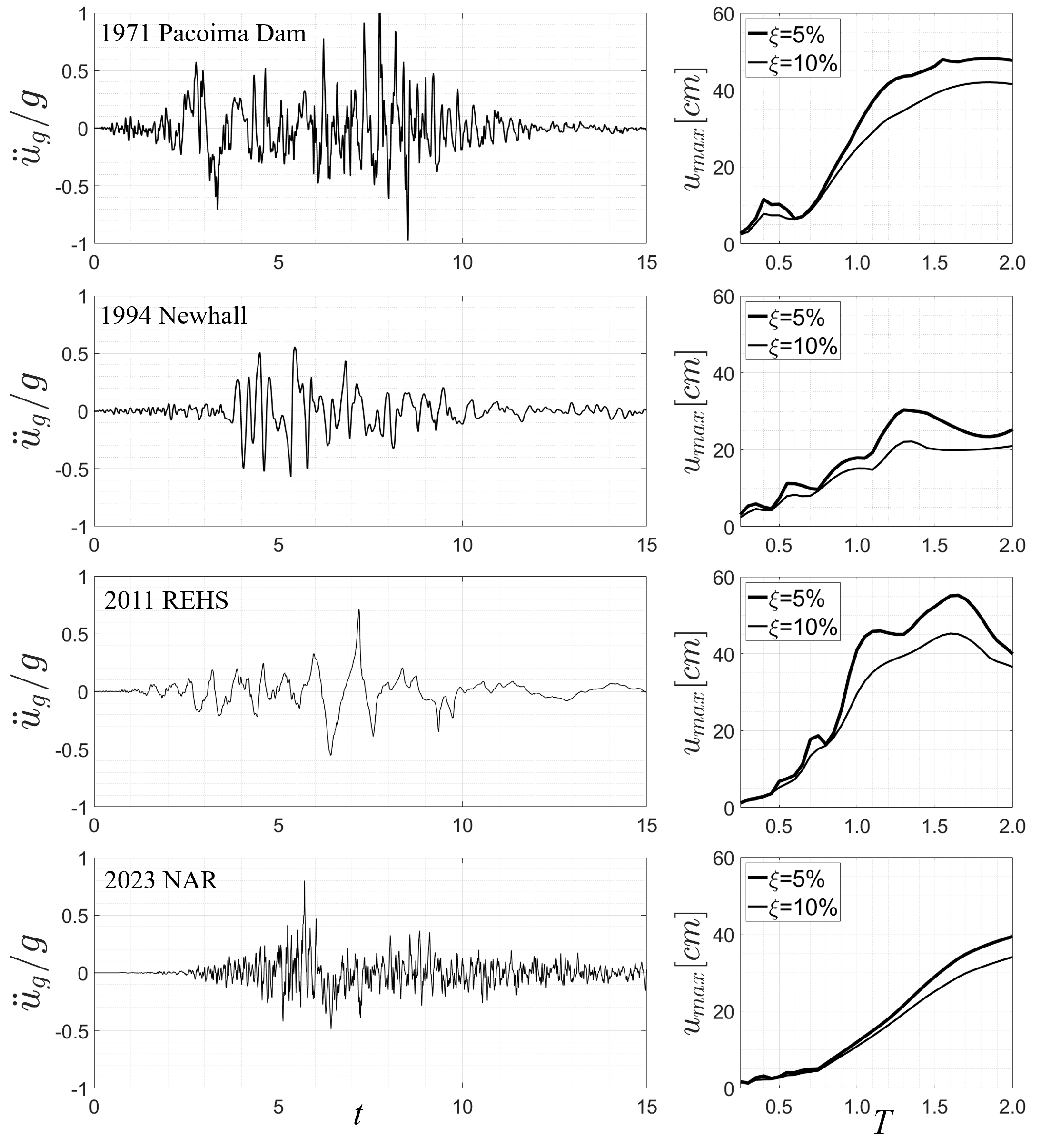}
\caption{Recorded time histories and elastic response spectra for damping ratio $\xi=\frac{c}{2m\omega_o}=5\%$ and $10\%$ of the four ground motions used for the response analysis presented in this study.}
\label{fig:ground_motions}
\end{figure}

For the 9-story MRF as it is shown in Figure (\ref{fig:segmental_wall}-left) a shear wall with height of $37.17$~m and width of $6.20$~m is selected (width/height = $\tan\alpha \approx$ $1/6$). For the case of 20-story building (shown in Figure (\ref{fig:segmental_wall}-right)) a shear wall with height of $80.73$~m and width of $8.0$~m is selected (width/height = $\tan\alpha \approx$ $1/8$).

\begin{figure}[t!]
\centering
\includegraphics[width=\textwidth]{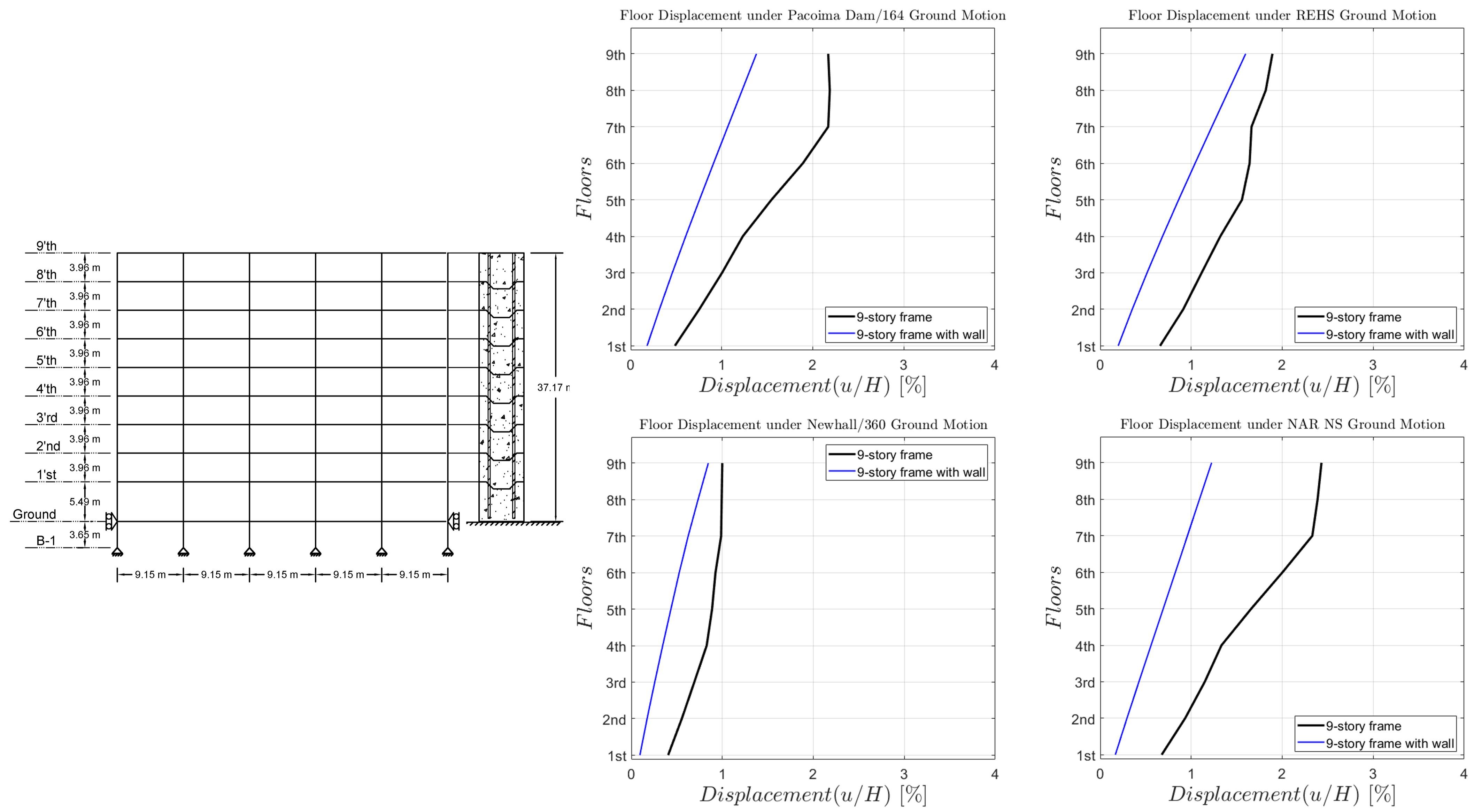}
\caption{Maximum floor displacement (normalized with respect to the total height of the wall the building) of the 9-story MRF without wall (dark heavy line in black) and the MRF with a stepping rocking wall (light line in blue) under various earthquake excitations.}
\label{fig:9_story_flr_disp}
\end{figure}

Dynamic time-history analysis of both 9-story and 20-story frames with and without wall is investigated in this section. This analysis is conducted applying various seismic records. Figure (\ref{fig:ground_motions}) presents recorded time histories and elastic response spectra for damping ratio $\xi=\frac{c}{2m\omega_o}=5\%$ and $10\%$. The ground motions used are the PacoimaDam/164 ground-motion recorded during the 1971 Imperial valley, the Newhall/360 ground motion recorded during the 1994 Northridge, the REHS ground motion recorded during the 2011 Christchurch, New Zealand, and the North-South component of NAR station ground motion recorded during 2023 Kahramanmaras, Turkiye earthquakes. The selected ground-motions were based on the reasoning that the distinguishable coherent pulse of these motions has different duration; therefore, each motion will amplify the inelastic structural response at different preyielding periods.

\begin{figure}[t!]
\centering
\includegraphics[width=\textwidth]{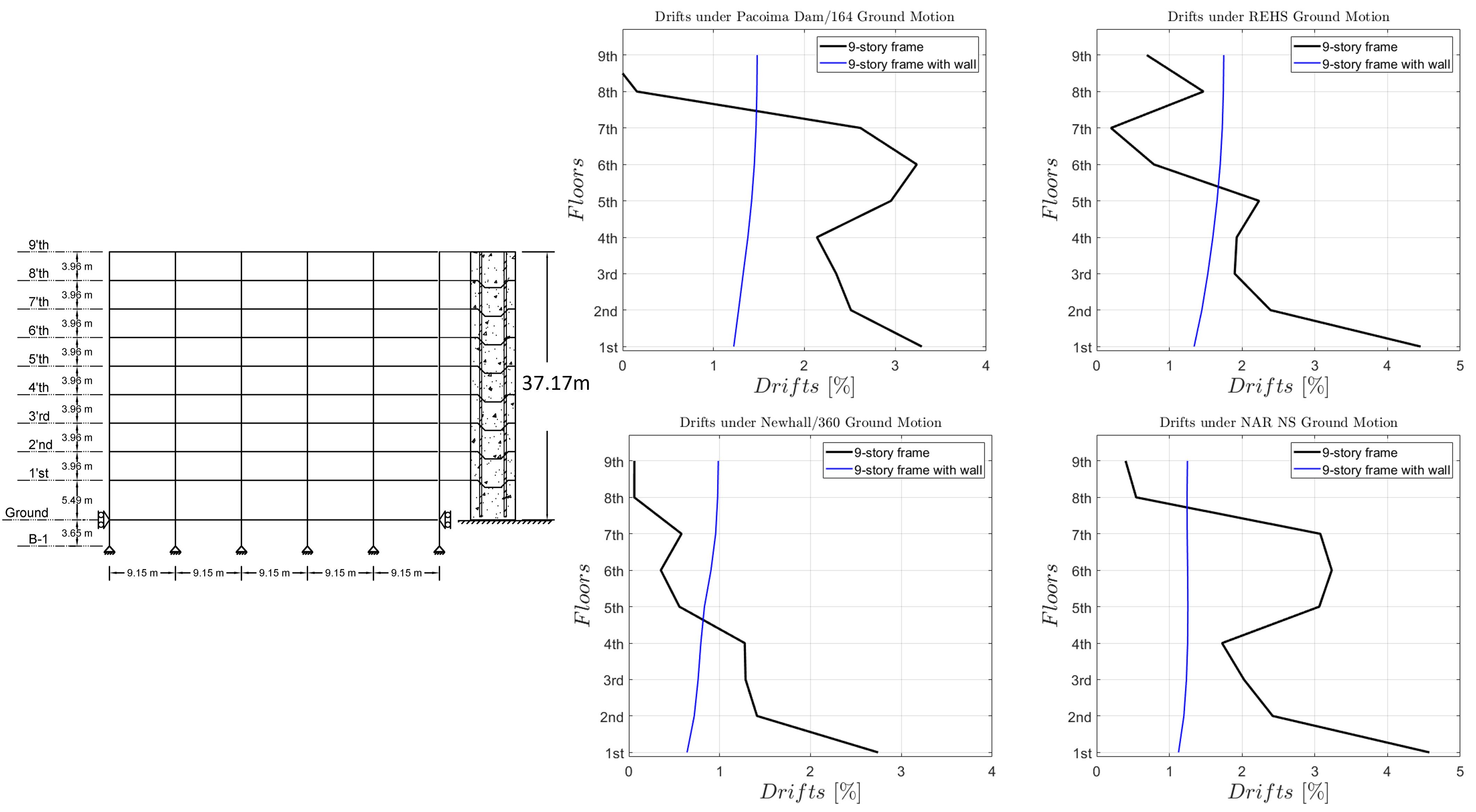}
\caption{Interstory drifts for the 9-story MRF without wall (dark heavy line in black) and the MRF with a stepping rocking wall (light line in blue) under various earthquake excitations.}
\label{fig:9_story_drift}
\end{figure}

Figures (\ref{fig:9_story_flr_disp}) and (\ref{fig:9_story_drift}) show the maximum floor displacement, that is normalized with respect to the above the ground height of the building, and interstory drifts for the 9-story MRF without wall (dark heavy line in black) and the MRF with a stepping rocking wall (light line in blue) respectively. Results in the Figure (\ref{fig:9_story_flr_disp}) shows that for all the cases adding a stepping rocking wall suppress the maximum floor displacement. In addition, it shows that the the rocking wall enforces the floor displacements to be mostly first-mode dominant. Most importantly, the main motivation of using rocking walls is to prevent weak-story failure. In order to investigate this, it is important to check the interstory drifts. Figure (\ref{fig:9_story_drift}) shows the comparison of the interstory drifts under various earthquakes. The results show that the addition of rocking wall creates mostly a uniform drift profile when it is compared to the interstory drift profile of the frame without a wall. This will prevent generation of weak-stories since the displacements are distributed evenly along the height of the building.

\begin{figure}[b!]
\centering
\includegraphics[width=\textwidth]{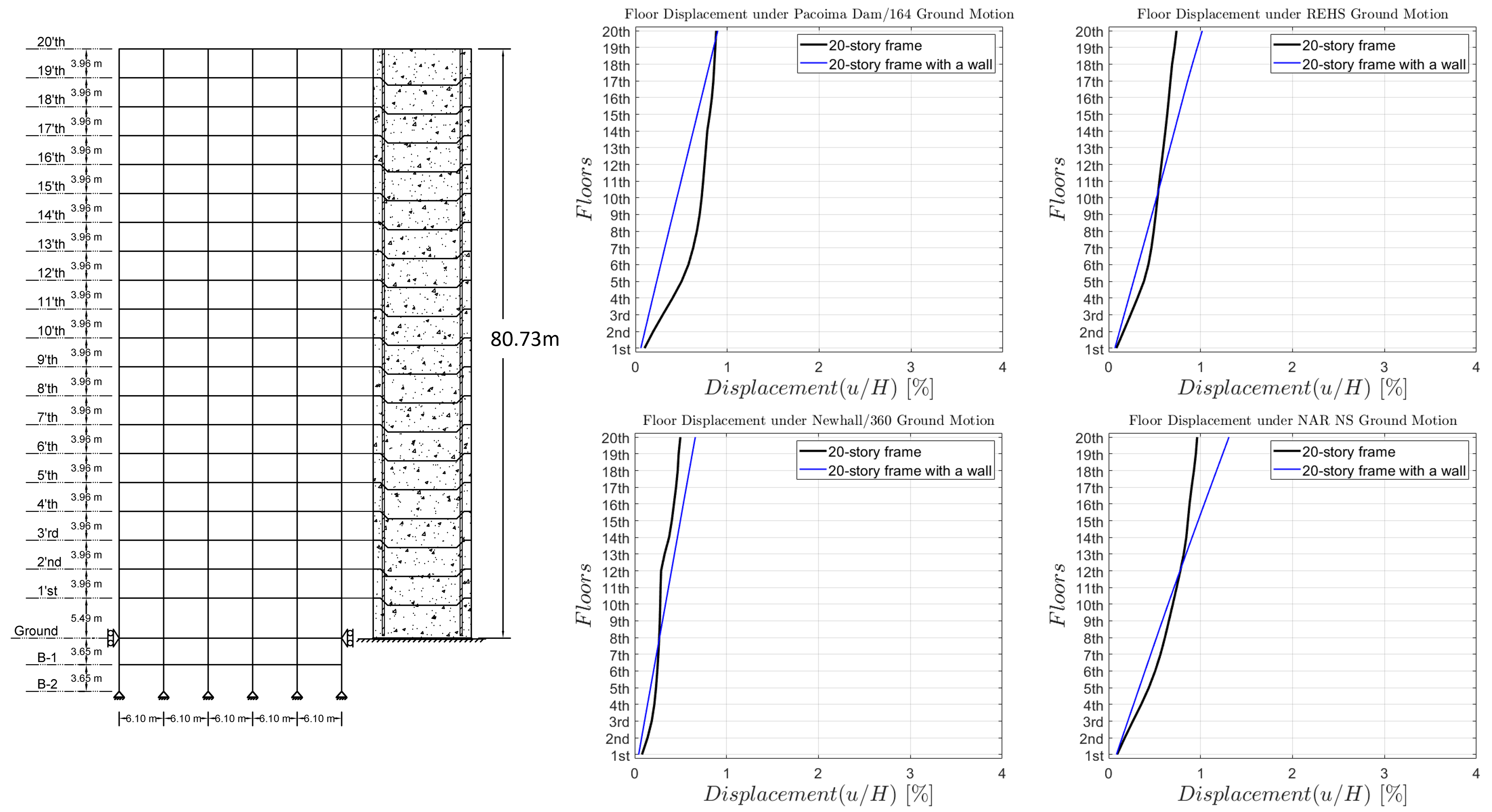}
\caption{Maximum floor displacement (normalized with respect to the total height of the wall the building) of the 20-story MRF without wall (dark heavy line in black) and the MRF with a stepping rocking wall (light line in blue) under various earthquake excitations.}
\label{fig:20_story_flr_disp}
\end{figure}

\begin{figure}[t!]
\centering
\includegraphics[width=\textwidth]{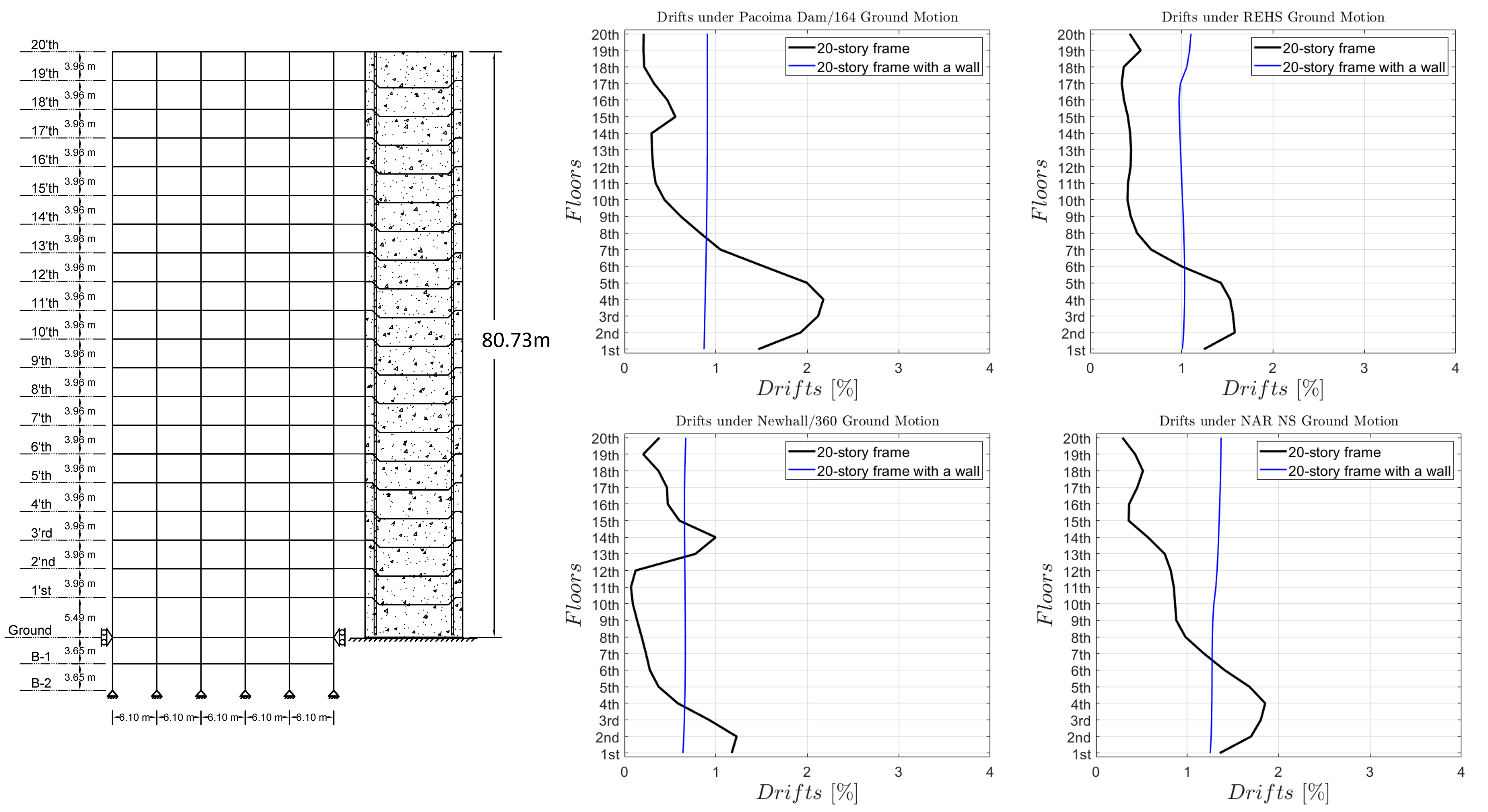}
\caption{Interstory drifts for the 20-story MRF without wall (dark heavy line in black) and the MRF with a stepping rocking wall (light line in blue) under various earthquake excitations.}
\label{fig:20_story_drift}
\end{figure}

Similarly, Figures (\ref{fig:20_story_flr_disp}) and (\ref{fig:20_story_drift}) show the maximum floor displacement, that is normalized with respect to the above the ground height of the building, and interstory drifts for the 20-story MRF without wall (dark heavy line in black) and the MRF with a stepping rocking wall (light line in blue) respectively. Floor displacement results (Figure (\ref{fig:20_story_flr_disp})) shows higher mode effects in the displacement of the 20-story frame when no wall is added. On the other hand, displacement profile of a 20-story MRF when it is coupled with a rocking wall is similar to first mode vibration of the building. Since the displacement profile for the frame with a wall is similar to the first-mode vibration, the interstory profile results, shown in the Figure (\ref{fig:20_story_drift}) is uniform along the height of the building. This is important since the main reason for the generation of weak-story is the presence of significant difference in interstory drifts that is visible for the 20-story frame without a shear wall.

\section{Conclusions}
The high occupancy rates in urban multi-story buildings, combined with present safety concerns, necessarily prompt a reassessment of performance goals. Given the notable seismic damage and instances of weak-story failures that have been documented after major earthquakes, this paper studied the use of modular shear walls that are free to rock above their foundation. This paper first conducted an overview of available literature on rocking elements and provided nonlinear equations of motions for spring-rocking-wall system for different configurations. Then a dynamic analysis for two 9-story and 20-story moment-resisting steel frames when they are coupled with modular rocking shear walls were investigated to reach the following conclusions. 

\begin{itemize}

\item The self-weight of the pinned rocking wall works against its stability. Hence, the stepping rocking wall will be more effective in reducing the residual displacement of the structure.

\item Time-history analysis of idealized SDOF nonlinear spring with and without rocking wall shows that the use of stepping rocking wall can effectively minimize the residual drifts after earthquakes.

\item The effect of the vertical tendons even when they are stiff $\frac{EA}{m_wg}=200$ and highly prestressed ($P_o=m_wg$) is marginal.

\item The vertical prestressing tendons will increase the vertical reactions at the pivoting corners. Hence, when they are employed, it is important to further investigate the effect of additional forces in the pivoting corners and retrofit the pivoting points to prevent generation of excessive damages.

\item Addition of a modular stepping rocking wall to a moment-resisting frame not only reduces its maximum floor displacement for most cases, it also enforces a first-mode dominating response. Hence prevents development of weak-story failure.

\end{itemize}

\section*{Declaration of Competing Interest}
The authors declare that they have no known competing financial interests or personal relationships that could have appeared to influence the work reported in this paper.

\section*{CRediT authorship contribution statement}
\textbf{Mehrdad Aghagholizadeh:} Conceptualization, Formal analysis, Investigation, Methodology, Software, Validation, Visualization, Writing - review \& editing.

\setstretch{.75}
\bibliography{references}
\end{document}